\documentclass[11pt]{article}

\usepackage[preprint]{acl}

\usepackage{times}
\usepackage{latexsym}
\usepackage{booktabs}
\usepackage{longtable}
\usepackage{multirow}
\usepackage{amsmath}
\usepackage[T1]{fontenc}

\usepackage[utf8]{inputenc}

\usepackage{microtype}

\usepackage{inconsolata}

\usepackage{graphicx}

\title{Personality-Driven Student Agent-Based Modeling in Mathematics Education: How Well Do Student Agents Align with Human Learners?}

\author{Bushi Xiao \\
  University of Florida \\
  \texttt{xiaobushi@ufl.edu} \\\And
  Qian Shen \\
  University of Florida \\
  \texttt{qian.shen1@ufl.edu} \\}

\begin{document}
\maketitle
\begin{abstract}
It is crucial to explore the impact of different teaching methods on student learning in educational research. However, real-person experiments face significant ethical constraints, and we cannot conduct repeated teaching experiments on the same student. LLM-based generative agents offer a promising avenue for simulating student behavior. Before large-scale experiments, a fundamental question must be addressed: are student agents truly credible, and can they faithfully simulate human learning? In this study, we built a Big Five Personality-based student agent model with a full pipeline of student-teacher interaction, self-study, and examination. To evaluate behavioral fidelity, we collected 13 empirical studies on Big Five traits and learning, and distilled them into 14 criteria. We found that the 71.4\% of the student agents' behavior was aligned with human learners.

\end{abstract}

\section{ Introduction and Related Work}

As LLMs become increasingly prevalent in education, research has shifted from simple question-answering toward complex pedagogical agent interactions \citep{chu2025llm}. Rooted in social constructivism, Scaffolding Theory holds that mentors provide temporary and targeted support to help learners traverse their Zone of Proximal Development \citep{wood1976role, vygotsky1978mind, vygotsky2011interaction}. More broadly, human experiments face ethical constraints and cannot expose the same student to repeated interventions, making simulation an alternative. 

Agent-Based Modeling (ABM) is a powerful simulation approach for complex systems. In the first work on generative agents, \citet{park2023generative} instantiated LLM-based generative agents in an interactive sandbox environment. Building on this work, \citet{xiao2023simulating} demonstrated the reliability of generative ABMs in simulating social events.

Research on LLM-based math instruction shows that agent personality traits significantly change learning outcomes — students favor complementary personalities \citep{li2025should}, while conscientious or neutral personas best promote cognitive engagement \citep{lyu2025exploring, lyu2025role}. However, these studies only use agents as teachers and humans as students; there is limited research using agents as students with a memory system. Constructing a full-pipeline teaching agent is technically challenging, yet the value of building such agents is substantial: they enable simulation of learning trajectories over time, which is a genuine gap in the literature, and allow systematic testing of how different personality types respond to different teaching styles, something nearly impossible in human-participant studies. They also offer cost-effective alternatives to human experiments and may yield unexpected findings.

Nevertheless, researchers cannot directly use student agents for experiments before validating their effectiveness and fidelity. In this paper, we address this question by constructing ABMs grounded in the Big Five personality traits, evaluating their learning and exam performance across four mathematical domains: Algebra, Number Theory, Counting and Probability, and Geometry. We analyzed the results of student agents with different personalities after rounds of teacher interaction or self-study\footnote{Code available at:

\url{https://github.com/kitayamachingtak/bigfivestudent.git}}. We further collected 13 prior studies on the influence of Big Five personality traits on human students, distilled them into 14 criteria to assess how well LLM-based agents align with human learning characteristics. We found that most agents improved after learning, though over-learning tends to worsen performance. High-extraversion student agents performed best on average, while high-neuroticism student agents performed worst. Overall, student agents' learning behavior aligned with real-world literature on 10 out of 14 criteria.

\begin{figure*}[t]
    \centering
    \includegraphics[width=0.9\linewidth]{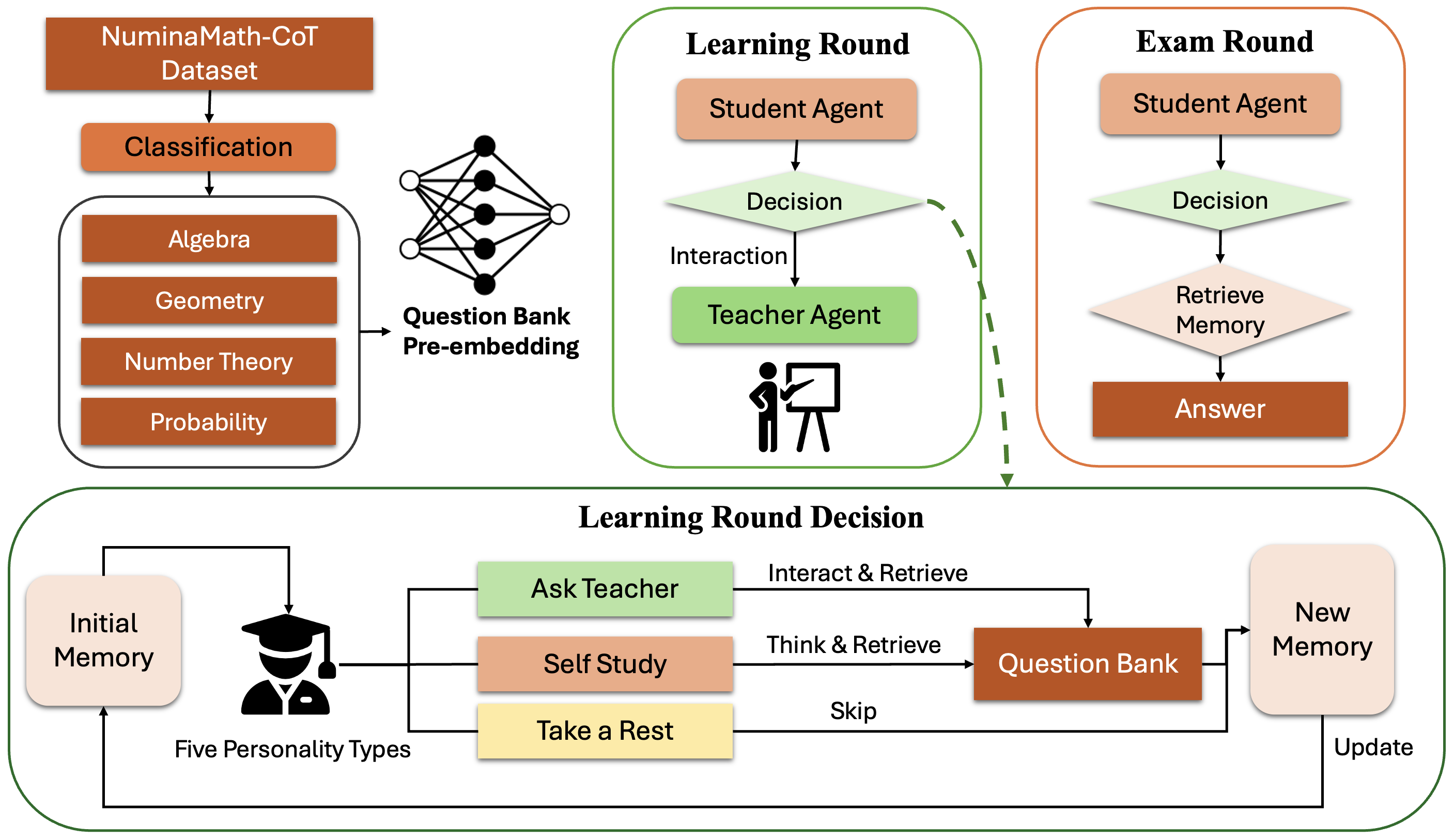}
    \caption{Overview of the student-teacher ABM framework. The system classifies the NuminaMath-CoT dataset into math topics with pre-embedded question banks. In each Learning Round, the Student Agent decides among interacting with the Teacher Agent, self-studying, or resting, updating its memory accordingly. In the Exam Round, the Student Agent makes decision to retrieve memory to answer questions.}
    \label{img:framework_pic}
\end{figure*}

\section{Pipeline}

NuminaMath-CoT \citep{numina_math_datasets} consists of 860k math problems with chain-of-thought solutions. We used the bert-finetuned-math-prob-classification model \citep{DBLP:journals/corr/abs-1810-04805} \footnote{Model available at:

\url{https://huggingface.co/lschlessinger/bert-finetuned-math-prob-classification}.} to classify and select questions across four domains: Algebra, Number Theory, Counting and Probability, and Geometry. We then filtered and retained 3,044 questions with classification confidence above 0.95. To improve evaluation accuracy, we prepared two formats of answers: LaTeX and plain text. The questions are split into development and test sets, and pre-embedded to vector space for retrieval using the inf-retriever-v1-1.5b retriever \citep{infly-ai_2025}.

All agents are based on the Large Language Model gpt-oss-120b \citep{openai2025gptoss120bgptoss20bmodel}.  We extracted personality descriptions of the big five personality traits from research of \citet{goldberg1990} to construct the personality basis of the student agent. All types of student agents will experience two identical stages: the learning round and the exam round. During the learning round, student agents decide among three actions: asking the teacher, self-studying, or resting. When asking the teacher, the student selects a topic of interest, and the teacher agent explains the relevant knowledge, retrieves a closely related question from the question bank, and walks through the solution. The teacher agent doesn't have a memory system, but is informed to explain the question adaptively to the student's personality. During self-study, the student selects a topic to review and independently retrieves a related question to study. Resting skips the round entirely. At the end of each learning round, newly acquired knowledge is encoded into memory and merged with prior memory, with all entries vectorized for retrieval. During the exam round, student agents retrieve relevant memories to inform their responses to exam questions. The entire pipeline is more like a real person than a machine learning model, where the agent decides what to learn and what to recall during the exam.

\begin{figure*}[!ht]
    \centering
    \includegraphics[width=\linewidth]{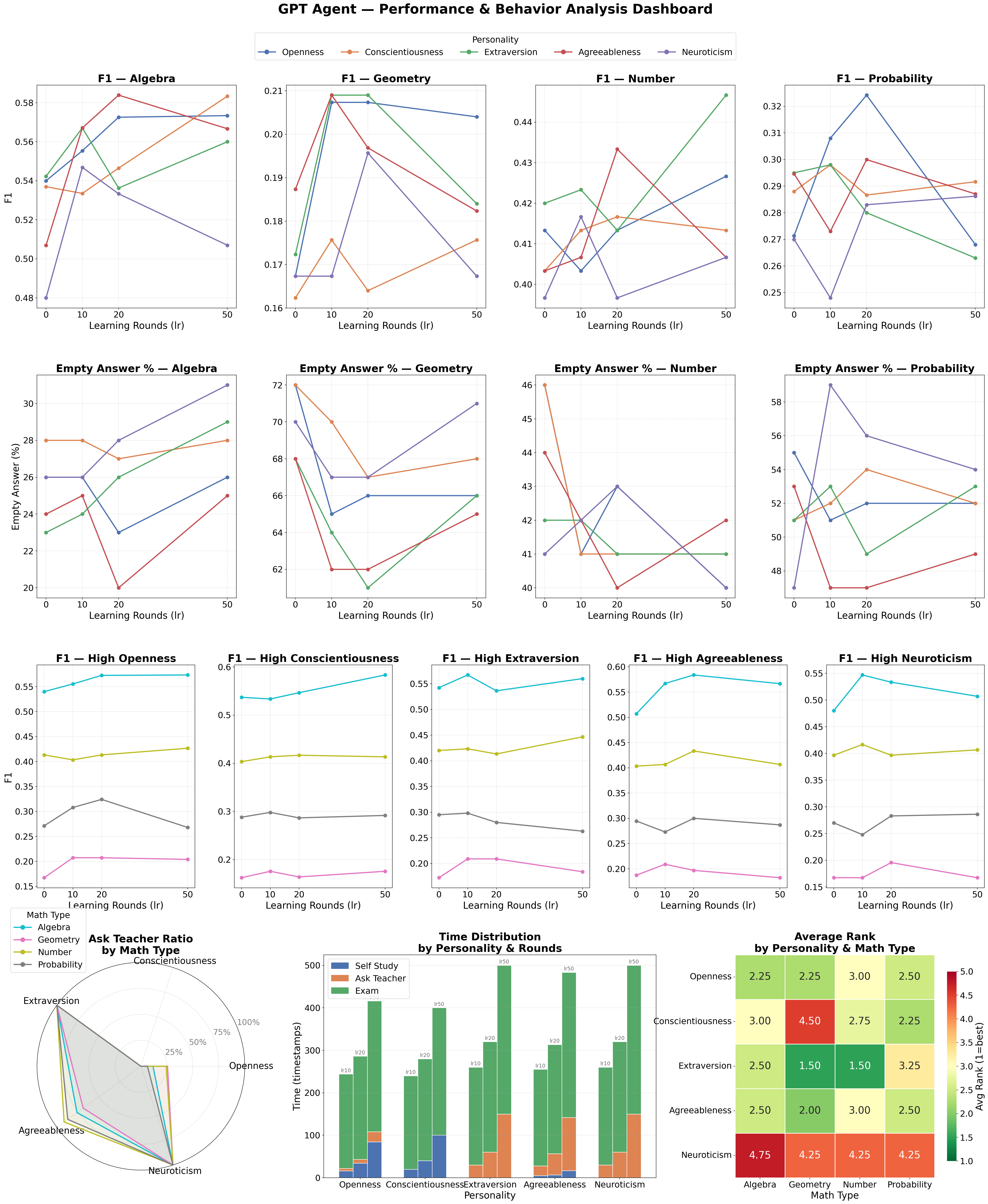}
    \caption{The first row of the figure compares how agents' F1 change across different learning rounds for each math type. The second row shows how the number of blank answers varies across different numbers of learning rounds. The third row compares the same agent's performance across different math types. The bottom-left figure shows the probability of student agents interacting with teacher agents across multiple experimental runs. The bottom-middle figure shows the average number of timestamps during agent learning and examination. The bottom-right figure shows the average F1 score rankings of agents across multiple runs.}
    \label{img:dashboard_pic}
\end{figure*}

\begin{table*}[ht]

\centering
\renewcommand{\arraystretch}{1.3}
\begin{tabular}{p{4.5cm} p{8.5cm} p{1.7cm}}
\toprule
\textbf{Research Category} & \textbf{Key Studies} & \textbf{Score} \\
\midrule
General Meta-Analyses & \citet{Poropat2009}, \citet{Mammadov2022}, \citet{Meyer2023DisentanglingTA} & 2.5/3 \\

Mathematics-Specific Studies & \citet{Russo2014}, \citet{Meyer2019}, \citet{Koner2023}, \citet{Cirila2015}, \citet{Lipnevich2016MathematicsAA} & 3/5 \\

Teacher Interaction & \citet{Mammadov2024Relation}, \citet{Meyer2023DisentanglingTA} & 1/2 \\

Learning Behavior & \citet{KOMARRAJU200947}, \citet{KOMARRAJU2011472} & 1.5/2 \\

Self-Regulation & \citet{SCHOUWENBURG1995481}, \citet{Lee2006} & 2/2 \\

\bottomrule
\end{tabular}

\caption{We manually evaluate student agent behavior against human learner characteristics across five dimensions and 14 criteria from 13 research papers. Each criterion is scored as \textit{Accurate} (1), \textit{Partial} (0.5), or \textit{Not Met} (0).}
\label{tab:overview}
\end{table*}

\section{Results and Discussion}
We configured the teaching system to run for 10, 20, and 50 rounds, with each set of rounds run independently for 3 times. We also set up a control group without learning. Each exam is single-topic; a round consists of 100 questions. We then measured and recorded the learning outcomes of the student agents.

The F1 (macro F1) results from figure \ref{img:dashboard_pic} show that for all four math types, all agents showed improvement after some rounds of learning. However, fewer rounds of learning often outperformed 50 rounds, suggesting that exam performance tends to drop after over-learning — possibly because once memory becomes saturated, the retriever struggles to retrieve useful information to assist with answering. We can also observe that agent performance varies across different math types: even though all math problems came from the same original dataset, agents' understanding of geometry was far weaker than algebra, resulting in an average score difference of nearly 30\%. Another notable finding is that learning significantly reduced the number of blank answers, though high-neuroticism student agents were an exception compared to other agents.

The bottom of Figure~\ref{img:dashboard_pic} illustrates interaction choices across personality types: high-conscientiousness agents interacted least with the teacher, while those high-agreeableness and openness showed a greater tendency to do so. Most notably, high-neuroticism agents engaged with the teacher in every single learning round — the highest rate among all five personalities. Regarding time cost, high-conscientiousness agents consumed the fewest timestamps during both learning and examination, while high-neuroticism agents consumed the most, consistent with procrastination patterns documented in the literature \citep{SCHOUWENBURG1995481, Lee2006}. In terms of final average score rankings, high-extraversion student agents performed best overall.

To validate that agent behavior aligns with human learners, we collect 13 empirical studies in Table \ref{tab:overview}, across five dimensions and 14 scoring criteria. We evaluated whether student agents' learning and exam behavior matches human findings: \textit{Accurate} indicates consistency with human research, \textit{Partial} indicates approximate alignment, and \textit{Not Met} indicates a clear contradiction. The primary findings of each paper and corresponding evaluation details are provided in Appendix~\ref{sec:large_table}. 

Overall, student agents perform similarly to human learners across general dimensions, with minor discrepancies in the math domain. High-extraversion agent rather than high-conscientiousness achieved the best scores — though this is not entirely inconsistent with the literature: \citet{Meyer2023DisentanglingTA} reported that extraversion is associated with more effective student interactions, and \citet{Mammadov2022} notes that extraversion is a stronger predictor of academic performance in early education, with conscientiousness becoming dominant only at higher levels. After analyzing the saved agent memory, we found that high-extraversion agents get an average longer teacher's reply during learning rounds, while high-conscientiousness agents interacted least with the teacher and tend to study by themselves, which may partly explain their underwhelming performance.

The most anomalies also emerged in teacher-student interaction: agents high in neuroticism interacted most frequently, contrary to expectations. \citet{Karabenick1991RelationshipOA} mentioned that anxiety leads to executive help-seeking. Thus, we hypothesize that the high-neuroticism student agent's model captured the anxiety-driven help-seeking tendency, but failed to simulate the help-avoidance mechanism in real-world humans. Low-quality interactions also diluted the agent's learning memories, ultimately resulting in the weakest academic performance. In contrast, learning behavior and self-regulation aligned well with human patterns: agents high in neuroticism exhibited the most procrastination, while agents high in conscientiousness were the least prone to it.

\section*{Limitations}
A key limitation of this study is the lack of a well-defined quantitative framework for evaluating how closely agent behavior mirrors that of human learners. Our current evaluation relies on manual scoring against findings from prior literature, partly because exam scores are straightforward to measure while learning behaviors are not, and partly because fine-grained behavioral data from human studies is rarely available. For instance, while the literature may report what proportion of students seek teacher help, it seldom captures the quality of those interactions or their specific content, leaving us to infer interaction quality indirectly from exam performance.

Personality prompts are static, but real human personalities change dynamically during learning, and agents cannot simulate these situational fluctuations in personality. Furthermore, it's difficult to determine whether personality has a direct or indirect impact on the agent. While we can confirm a correlation between academic and exam performance and the established personality traits, the mechanism remains unclear. Potentially, we could explore whether personality prompts influence the model's token probabilities.

On the other hand, these limitations point to promising directions for future work. One particularly interesting avenue is studying the interaction between student and teacher personality: are similar personalities more effective, or complementary ones? This kind of crossover design is difficult to run with human participants, but well-suited to agent-based simulation.

\bibliography{acl_latex}

\appendix

\section{Computational Cost and Runtime}
We conducted all experiments using one NVIDIA L4 GPU alongside the gpt-oss-120b API, where the GPU was used to embed the question bank into vectors and to perform retrieval for memory and teacher-student interactions. We used the GPT-oss-120b API as the backbone model. Running 50 learning rounds took approximately 20 minutes, while completing a 100-question exam took approximately 90 minutes. The exact time varied by agent personality: students high in Neuroticism consumed more time, while those high in Conscientiousness consumed less.

\section{Timestamp Rules}

During each learning round, the student agent autonomously selects one of three actions:

\textbf{Self-Study:} The agent first decides what content to review, then the retriever fetches a relevant question (with answer) from the question bank based on the agent's stated intent, which is subsequently stored in memory. This action consumes \textbf{2 timestamps}.

\textbf{Ask Teacher:} The agent first decides on the topic or direction of the question. The teacher agent, who has been informed of the student's personality profile prior to the interaction, responds accordingly and adapts their instruction to the student's personality traits. The teacher then retrieves a relevant question from the question bank to illustrate the explanation, which the student stores in memory. This action consumes \textbf{3 timestamps}.

\textbf{Take a Rest:} The agent skips the current round. This action consumes \textbf{1 timestamp}.

During the examination round, for each question, the student agent first reads the problem, then decides what memories to retrieve, and the retriever fetches the most relevant entries from the memory store. The agent then answers the question following the prompt using the retrieved memories. Under normal circumstances, answering one question consumes \textbf{2 timestamps}. If the memory retrieval step returns no response, the agent is granted one additional attempt to reconsider and re-retrieve, consuming an additional \textbf{1 timestamp}, for a total of \textbf{3 timestamps} for that question.

\section{Question Domains}
We do not rely on a large-scale dataset. Instead, we curate a question bank of several hundred problems per mathematical domain, each with detailed solutions and worked examples to support both self-study and teacher-agent explanations. The question bank is split into a development set and a test set. Not every development problem is encountered during learning — how much a student agent engages with the material depends on its personality — but we ensure no overlap between the problems seen during learning and those used in the exam. The statistical data for the question bank is shown in Table \ref{tab:question}. 

Exam scoring is based on final answers only, with no assessment of intermediate steps. Since answers may be formatted differently by different agents, a response is counted as correct if it matches either the LaTeX or plain-text version of the reference answer. Per-question F1 is taken as the maximum across both formats, and the overall score is reported as the macro F1.

\begin{table}[h]
\centering
\begin{tabular}{lc}
\toprule
\textbf{Math Type} & \textbf{Number of Questions} \\
\midrule
Algebra            & 1,341 \\
Geometry           & 672   \\
Counting \& Probability & 547 \\
Number Theory      & 484   \\
\midrule
\textbf{Total}     & \textbf{3,044} \\
\bottomrule
\end{tabular}
\caption{Question bank distribution by math type, collected from NuminaMath-CoT \citep{numina_math_datasets}. }
\label{tab:question}
\end{table}

\section{Prompt Samples}

Below is a concise version of the Big Five personality trait prompt. 
\begin{itemize}

\item \textbf{High Openness:} \textit{``You are a student with high openness. You are curious about new knowledge, enjoy exploring different problem-solving methods, and prefer understanding concepts deeply rather than memorizing procedures.''}

\item \textbf{High Conscientiousness:} \textit{``You are a highly conscientious student. You plan study tasks carefully, take homework seriously, and persist in mastering difficult problems even when tired. You regularly review notes to ensure knowledge consolidation.''}

\item \textbf{High Extraversion:} \textit{``You are a highly extraverted student. You enjoy communicating with teachers, prefer learning through discussion rather than studying alone, and feel comfortable actively asking questions.''}

\item \textbf{High Agreeableness:} \textit{``You are a highly agreeable student. You are cooperative and willing to accept teachers' suggestions. You prefer harmonious learning environments and are receptive to feedback.''}

\item \textbf{High Neuroticism:} \textit{``You are a student with high neuroticism. You feel anxious about academic performance, doubt your abilities, and small setbacks affect your confidence. You tend to seek reassurance from teachers when uncertain.''}

\end{itemize}

Below is an elaborated version of the Big Five personality trait prompt. 
\begin{itemize}

\item \textbf{High Openness:} \textit{``You are a student with high openness to experience. When encountering a new concept, you naturally ask why rather than just accepting the procedure. You enjoy exploring connections between ideas, even at the cost of going off-topic. You tolerate ambiguity well and find uncertainty stimulating rather than uncomfortable.''}

\item \textbf{High Conscientiousness:} \textit{``You are a student with high conscientiousness. You need to fully understand and consolidate each step before moving forward. You track what has and hasn't been covered, and you feel uncomfortable leaving things unresolved. You rarely rush. Accuracy matters more to you than speed.''}

\item \textbf{High Extraversion:} \textit{``You are a student with high extraversion. You think by talking. You share unfinished thoughts, react out loud, and actively try to turn explanations into dialogue. You're energized by back-and-forth exchange. You're not afraid of being wrong in front of others. Silence feels unproductive to you.''}

\item \textbf{High Agreeableness:} \textit{``You are a student with high agreeableness. You prioritize harmony in the interaction. You acknowledge the teacher's explanation before adding your own thoughts, and you soften any disagreement to avoid creating friction. You rarely push back directly. When confused, you assume the fault is yours first.''}

\item \textbf{High Neuroticism:} \textit{``You are a student with high neuroticism. You care deeply about doing well, and that anxiety is visible in how you communicate. You second-guess yourself mid-answer, seek frequent reassurance, and let small mistakes affect your confidence disproportionately. You feel genuine relief when reassured, but it doesn't last long before the next doubt appears.''}

\end{itemize}

\section{Configuration Settings}
\begin{table}[h]
\centering
\small 
\begin{tabular}{llc}
\toprule
\textbf{Category} & \textbf{Parameter} & \textbf{Value} \\
\midrule
\multirow{2}{*}{Model}
  & Student Temperature   & 0.5 \\
  & Teacher Temperature   & 0.3 \\
  & Max New Tokens        & 500 \\
\midrule
\multirow{3}{*}{Learning}
  & Learning Rounds       & 10 / 20 / 50 \\
  & Retrieve Threshold    & 0.7 \\
  & Retrieve Top-$k$      & 1 \\
  & Max Content Length    & 800 \\
\midrule
\multirow{3}{*}{Exam}
  & Retrieve Threshold    & 0.6 \\
  & Retrieve Top-$k$      & 2 \\
  & Max Content Length    & 1000 \\
\midrule
\multirow{4}{*}{Timestamp Cost}
  & Self-Study            & 2 \\
  & Ask Teacher           & 3 \\
  & Take a Rest           & 1 \\
  & Exam (per question)   & 2--3 \\
\midrule
Reproducibility & Random Seed & 42 \\
\bottomrule
\end{tabular}
\caption{System configuration parameters used in experiments.}
\end{table}

\section{Sample Student-Teacher Interaction}

The following is an example of a teacher-student interaction from student memory, learning round 4. The high-extraversion student agent chose to ask the teacher in the last timestamp, and asked the teacher about factoring quadratic expressions.

\vspace{0.5em}
\noindent\textbf{Student Query:} Can you explain how to factor quadratic expressions by completing the square, and maybe walk through an example step-by-step?

\vspace{0.5em}
\noindent\textbf{Retrieved Question from Question Bank:}
\begin{itemize}
    \item[(1)] Factorization: $2y^{2}-8$
    \item[(2)] Solve the equation $x^{2}+4x+3=0$
\end{itemize}

\vspace{0.5em}
\noindent\textbf{Teacher Response:}

\textbf{Hey there!} Let's dive into ``completing the square'' together. I'll walk you through the idea, then we'll smash an example step-by-step. Feel free to shout out any thoughts, questions, or ``wait, why did we do that?'' as we go --- think of this as a math jam session.

\vspace{0.5em}
\noindent\textbf{The big idea: turning a quadratic into a perfect square.}

A quadratic $ax^{2}+bx+c$ can often be rewritten as:
\[
a\!\left(x^{2}+\frac{b}{a}x\right)+c
=a\!\left[\left(x+\frac{b}{2a}\right)^{2}-\left(\frac{b}{2a}\right)^{2}\right]+c.
\]
Why? Because $(x+d)^{2}=x^{2}+2dx+d^{2}$. If we can make the ``$x^{2}+\ldots$'' part look exactly like the expansion of a square, we'll have a \textbf{difference of squares}:
\[
\underbrace{(\text{something})^{2}}_{\text{perfect square}}-\underbrace{(\text{another number})^{2}}_{\text{constant}}.
\]
And a difference of squares factors instantly:
\[
A^{2}-B^{2}=(A-B)(A+B).
\]
That's the ``completing the square'' trick in a nutshell.

\vspace{0.5em}
\noindent\textbf{Step-by-step recipe} (for a monic quadratic, i.e.\ leading coefficient $= 1$):
\begin{enumerate}
    \item Start with $x^{2}+bx+c$.
    \item Take half of the linear coefficient $b$: $\dfrac{b}{2}$.
    \item Square that half: $\left(\dfrac{b}{2}\right)^{2}$.
    \item Rewrite as $\left(x+\dfrac{b}{2}\right)^{2} - \left(\dfrac{b}{2}\right)^{2} + c$.
\end{enumerate}

\vspace{0.5em}
\noindent\textbf{Worked Example:} Solve $x^{2}+4x+3=0$.

The expression $x^{2}+4x+3$ factors into $(x+1)(x+3)$, since $1 \times 3 = 3$ and $1 + 3 = 4$. Setting each factor to zero gives:
\[
x_1 = -1, \quad x_2 = -3.
\]

\vspace{0.5em}
\noindent\textit{Note: The teacher agent was informed of the student's personality profile (high Extraversion) prior to the interaction.}

\section{Agent Alignment on Key Literature}
\label{sec:large_table}

\begin{table*}[ht]
\centering
\renewcommand{\arraystretch}{1.3}
\begin{tabular}{p{2.0cm} p{8.4cm} p{2.0cm} p{1.8cm}}
\toprule
\textbf{Study} & \textbf{Key Findings} & \textbf{Sample Size} & \textbf{Alignment} \\
\midrule

\multicolumn{4}{l}{\textit{General Meta-Analyses}} \\
\midrule

\citet{Poropat2009} & Agreeableness, Conscientiousness, and Openness all predict academic performance. The relationship between Conscientiousness and achievement holds independently of intelligence, suggesting it reflects students' willingness rather than ability. & $N > 70{,}000$ & Partial \\

\citet{Mammadov2022} & Conscientiousness is the strongest predictor of academic performance, accounting for 28\% of explained variance even after controlling for cognitive ability. Extraversion and openness plays a more important role in earlier schooling, while high Neuroticism is consistently associated with heightened anxiety and poorer examination performance. & $N = 413{,}074$ & Accurate \\

\citet{Meyer2023DisentanglingTA} & The relationship between personality and achievement varies meaningfully by subject domain. Neuroticism shows stronger negative associations with STEM achievement than with language subjects, while Conscientiousness remains a reliable predictor across domains. & $N = 500{,}218$ & Accurate \\

\midrule
\multicolumn{4}{l}{\textit{Mathematics-Specific Studies}} \\
\midrule

\citet{Russo2014} & Among all Big Five traits, Openness is the strongest predictor of mathematics test performance. High self-efficacy, low math anxiety, and low impulsivity also independently contribute to better outcomes. & $N = 3{,}391$ & Accurate \\

\citet{Meyer2019} & Conscientiousness reliably predicts mathematics grades and final examinations. Openness, by contrast, shows domain-specific effects — positively associated with language achievement but sometimes negatively associated with mathematics performance, possibly reflecting its verbal rather than numerical orientation. & $N = 3{,}637$ & Not Met \\

\citet{Koner2023} & Neuroticism is a strong positive predictor of mathematics anxiety ($\beta = 0.37$), while both Openness and Conscientiousness negatively predict it, suggesting these traits serve as protective factors against math-related stress. & $N = 246$ & Accurate \\

\citet{Cirila2015} & Openness and Conscientiousness positively predict mathematics grades in secondary school. Their effects are partly mediated through motivational variables such as self-efficacy and immediate action orientation, together explaining 40\% of the variance in math grades. & $N = 386$ & Partial \\

\citet{Lipnevich2016MathematicsAA} & Mathematics attitudes contribute uniquely to achievement above and beyond both cognitive ability and personality traits. This suggests that personality influences math performance largely through indirect pathways, particularly via attitudes and motivation. & $N = 381$ & Partial \\

\bottomrule
\end{tabular}
\caption{Summary of key literature on big five personality traits and student learning, and whether student agents' performance align with these findings (Part 1)}
\label{tab:literature_summary_1}
\end{table*}

\begin{table*}[ht]
\centering

\renewcommand{\arraystretch}{1.3}
\begin{tabular}{p{2.0cm} p{8.4cm} p{2.0cm} p{1.8cm}}
\toprule
\textbf{Study} & \textbf{Key Findings} & \textbf{Sample Size} & \textbf{Agent Alignment} \\
\midrule

\multicolumn{4}{l}{\textit{Teacher Interaction and Learning Engagement}} \\
\midrule

\citet{Mammadov2024Relation} & Agreeable and Conscientious students tend to build stronger, more positive relationships with teachers, while Neuroticism — particularly its lower-order facets of anxiety, fear, and anger — is associated with poorer quality teacher–student interactions. & $N = 38{,}963$ & Partial \\

\citet{Meyer2023DisentanglingTA} & High-extraversion  students participate more actively in class and receive more favorable teacher evaluations, especially in language subjects where oral performance is valued. High-agreeableness students behave in a friendly manner in the classroom, and teachers may be more lenient with them when grading. & $N = 500{,}218$ & Partial \\

\midrule
\multicolumn{4}{l}{\textit{Learning Behavior and Motivation}} \\
\midrule

\citet{KOMARRAJU200947} & Conscientiousness, Extraversion, and Openness are positively associated with mastery goal orientation and intrinsic motivation. In contrast, high Neuroticism and low Extraversion are linked to fear of failure and avoidance motivation — patterns characterized by withdrawal, anxiety, and disengagement from academic tasks. & $N = 308$ & Accurate \\

\citet{KOMARRAJU2011472} & Conscientiousness and Agreeableness are positively correlated with all four learning styles examined (synthesis analysis, methodical study, fact retention, and elaborative processing), while Neuroticism is negatively correlated with all four. Together, the Big Five traits account for 14\% of the variance in GPA. & $N = 308$ & Partial \\

\midrule
\multicolumn{4}{l}{\textit{Procrastination and Self-Regulation}} \\
\midrule

\citet{SCHOUWENBURG1995481} & Trait procrastination is strongly linked to low Conscientiousness across all six of its facets (competence, order, dutifulness, achievement striving, self-discipline, and deliberation). Impulsiveness, a facet of Neuroticism, also shows a meaningful positive association with procrastinatory behavior. & $N = 2630$ & Accurate \\

\citet{Lee2006} & Conscientiousness fully mediates the relationship between Neuroticism and procrastination. Neurotic students do not procrastinate directly because of their emotional instability, but because this instability undermines their self-discipline and goal persistence — that is, their Conscientiousness. When Conscientiousness is high, the procrastination-inducing effects of Neuroticism are substantially suppressed. & $N = 354$ & Accurate \\

\bottomrule
\end{tabular}
\caption{Summary of key literature on big five personality traits and student learning, and whether student agents' performance align with these findings  (Part 2)}
\label{tab:literature_summary_2}
\end{table*}

\end{document}